\newcommand{\beq}{\begin{equation}}
\newcommand{\eeq}{\end{equation}}
\newcommand{\eq}[1]{\begin{align}#1\end{align}}
\newcommand{\lf}{\left}
\newcommand{\rf}{\right}
\newcommand{\nt}{\notag}
\newcommand{\beqa}{\begin{eqnarray}}
\newcommand{\eeqa}{\end{eqnarray}}

\documentclass[aps,prl,twocolumn,superscriptaddress]{revtex4-1}

\usepackage{graphicx}
\usepackage{verbatim}
\usepackage{mathrsfs}
\usepackage{cancel}
\usepackage{bm}
\usepackage{float}
\usepackage{color}
\usepackage{amsmath,amsfonts,amssymb}
\usepackage{hyperref}
\usepackage{cleveref}
\usepackage{epstopdf}

\begin{document}

\title{Valley Edelstein Effect in Monolayer Transition Metal Dichalcogenides}

\author{Katsuhisa Taguchi}
\affiliation{Department of Applied Physics, Nagoya University, Nagoya, 464-8603, Japan}

\author{Benjamin T. Zhou}
\affiliation{Department of Physics, Hong Kong University of Science and Technology, Clear Water Bay, Hong Kong, China}

\author{Yuki Kawaguchi}
\affiliation{Department of Applied Physics, Nagoya University, Nagoya, 464-8603, Japan}

\author{Yukio Tanaka}
\affiliation{Department of Applied Physics, Nagoya University, Nagoya, 464-8603, Japan}

\author{Kam Tuen Law}
\affiliation{Department of Physics, Hong Kong University of Science and Technology, Clear Water Bay, Hong Kong, China}

\begin {abstract}
 In this work, we predict the emergence of the valley Edelstein Effect (VEE), which is an electric-field-induced spin polarization effect, in gated monolayer transition metal dichalcogenides (MTMDs). We found an unconventional valley-dependent response in which the spin-polarization is parallel to the applied electric field with opposite spin-polarization generated by opposite valleys. This is in sharp contrast to the conventional Edelstein effect in which the induced spin-polarization is perpendicular to the applied electric field. We identify the origin of VEE as combined effects of conventional Edelstein effect and valley-dependent Berry curvatures induced by coexisting Rashba and Ising SOCs in gated MTMDs. Experimental schemes to detect the VEE are also considered.
\end{abstract}

\maketitle

\textbf{Introduction.}--
Monolayer transition metal dichalcogenides (MTMDs) have attracted much attention recently because of their peculiar electronic and optical properties \cite{Xu2014}. 
Semiconducting MTMDs, MX$_2$, are composed of transition metal atoms (M=Mo, W) and group-VI dichalcogenide atoms (X=S, Se, Te, etc.) \cite{Splendiani2010,Mak2010,Wang2012,Geim2013}.
They are arranged in two-dimensional (2D) honeycomb lattice structures
, and exhibit a direct band gap between the valence and conduction band edges near the $\pm K$ points \cite{Kuc2011,Zhu2011,Zahid2013,Shi2013,Cappelluti2013}. 
Both the top valence and the bottom conduction band edges of MTMDs are spin-split ($\sim 0.1 \ {\rm eV}$ and $\sim 10\ {\rm meV}$, respectively) due to strong atomic spin-orbit coupling (SOC) of the $d$-orbitals from transition metal atoms 
and in-plane mirror symmetry breaking
 in the lattice structure \cite{Yuan2013,Rostami2013,Kormanyos2013,Liu2013,Ochoa2013}.
In particular, the SOC here acts as a valley-dependent Zeeman field, called Ising SOC \cite{Lu2015,Xi2016}, which pins electron spins at opposite valleys to opposite out-of-plane directions.
Such a valley-dependent band structure makes MTMDs potential candidates for valleytronics devices \cite{Rose2013,Song2015,Xiao2012}. 
Several valley-dependent phenomena, such as valley-selective circularly dichroism \cite{Cao2012} and intrinsic valley Hall effect \cite{Mak2014}, have been theoretically studied and experimentally reported.
Besides, in gated MTMDs, superconductivity with nonzero Rashba SOCs and Ising SOCs have also been experimentally studied \cite{Lu2015,Saito2015}.

The relatively small Ising SOC in the conduction bands was ignored in previous studies \cite{Zahid2013, Shi2013, Cappelluti2013, Xiao2012}. In this Letter, we show that the valley-dependent Ising SOC together with the Rashba SOC generate strong Berry curvatures in the conduction bands. This Berry curvature combining with the conventional Edelstein effect in gated MTMDs leads to a new type of valley-dependent phenomenon, which we call the valley Edelstein effect (VEE). In conventional Edelstein effects\cite{Aronov1989,Edelstein1990,Chaplik2002,Inoue2002,Shen2014,Yoda2015}, the spin polarizations are generated by Rashba SOCs under an applied electric field $\bm{E}$
\cite{Kato2004,Silov2005,Isasa2016,Borge2014}, and the induced spin polarizations are perpendicular to $\bm{E}$. In the VEE, however, the induced spin polarization has an extra parallel component with respect to $\bm{E}$, with the polarizations generated by electrons from opposite valleys pointing to opposite directions.

\begin{figure}[b]
\centering
\includegraphics[width=8.6cm]{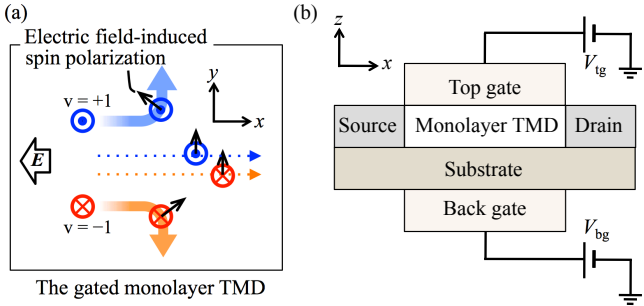}
\caption{\label{Fig1} 
Schematic of the VEE. The blue and orange arrows represent the motion of electrons in the ${\rm v}=+1$ and $-1$ valleys, respectively, induced by an electric field $\bm{E}$ applied in the $-x$ direction. The thick solid (thin dotted) arrows correspond to the trajectories in the presence (absence) of the Ising SOC. Due to valley-dependent Berry curvatures (Eq. (\ref{eq:BerryCurvature})) induced by Rashba and Ising SOCs, the trajectory of electrons from ${\rm v}=+1$ ($-1$) valley bends into the $+y$ ($-y$) direction. The spin polarization (black arrows) arises in the perpendicular direction to the electron motion via the spin-momentum locking due to the Rashba SOC. As a result, valley-dependent spin polarization is induced along $\bm{E}$ (VEE) in addition to the net spin polarization perpendicular to $\bm{E}$ (conventional Edelstein effect).
(b) Schematic dual-gate setup with tunable chemical potential and Rashba SOC. Top(back) gate voltage is indicated by $V_{\textrm{tg}}$($V_{\textrm{bg}}$). The chemical potential is tuned by the average of $V_{\textrm{tg}}$ and $V_{\textrm{bg}}$, while the Rashba SOC can be induced by the difference between $V_{\textrm{tg}}$ and $V_{\textrm{bg}}$.
}
\end{figure}

Remarkably, the unconventional parallel spin density calculated from Keldysh-Green's function method is proportional to the Berry curvature induced by the coexisting Rashba and Ising SOCs in gated MTMDs \cite{Lu2015, HYuan2014}. Physically, the Berry curvature drives electrons to drift in transverse directions under the applied electric field $\bm{E}$, and by combining with the conventional Edelstein effect spin components parallel to $\bm{E}$ can emerge [Fig. \ref{Fig1}(a)].

Importantly, the Berry curvature in VEEs results from a massive-Dirac-like Hamiltonian in spin basis [Eq. (\ref{eq:2-2})]\cite{SupplementaryTMD}. The Ising SOC plays the role of a Dirac mass term and has opposite signs at opposite valleys. This is very different from the intrinsic Berry curvature in pristine MTMDs studied previously \cite{Xiao2012}, in which valley-dependent Berry curvatures arises from orbital degrees of freedom.

To be specific, the spin density induced in response to $\bm E$ is given by
\beqa 
\label{eq:1-2} 
\langle \bm{s}^{\textrm{VEE}}_{\textrm{v}} \rangle 
	& = e \nu_e \left[\mathcal{C}_\perp (\hat{\bm{z}}\times \bm{E}) +\textrm{v}  \mathcal{C}_\parallel \bm{E}\right], 
\eeqa 
where $\textrm{v}=\pm$ is the valley index, $e<0$ is the electron charge, $\nu_e=m/(2\pi\hbar^2)$ is the 2D density of states, $m$ is the effective electron mass, $\hat{\bm z}$ is the unit vector normal to the 2D plane, and $\mathcal{C}_\perp$ and $\mathcal{C}_\parallel$ are the response coefficients for perpendicular and parallel spin components, respectively. The key finding of VEEs in this work is manifested in the non-zero value of $\mathcal{C}_{\parallel}$ in the second term of Eq. (\ref{eq:1-2}), which arises when both Rashba and Ising SOCs are present. 

In the following sections, we first present the model Hamiltonian for gated MTMDs (shown schematically in Fig. \ref{Fig1}(b)), which incorporates both impurity scattering effects and coupling to external electric fields. Second, we use Keldysh Green's function method to calculate the induced spin density within a linear response theory and show explicitly the emergence of the unconventional $\mathcal{C}_\parallel$. Remarkably, we demonstrate that $\mathcal{C}_\parallel$ is directly related to Berry curvatures induced by Rashba and Ising SOCs. Finally, we discuss experimental realization of VEE in gated MTMDs and explain how it can be detected by longitudinal magneto-optic Kerr effects \cite{You1998}.

\begin{figure}[b]
\includegraphics[width=8.5cm]{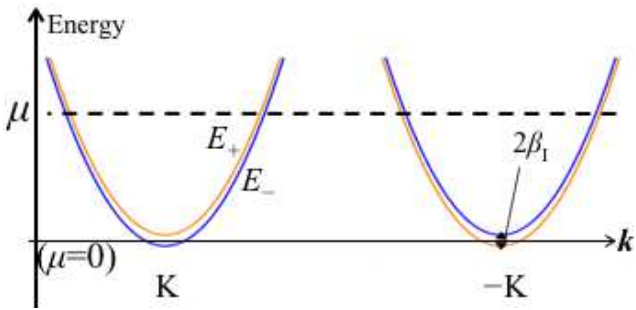}
\caption{ (Color online) 
Spin-split conduction bands of the MTMD with the Ising SOC and Rashba SOCs around $K$-point and -$K$-point   
 for $\mu  > \beta_{\textrm{I}}$.  
}
\label{fig:2} 
\end{figure}

\textbf{Model.}--
\label{sec:2} 
We consider a MTMD, such as MoS$_2$ and WS$_2$, and assume that the Fermi level crosses the spin-split conduction bands around the $\pm K$ points \cite{Wang2012,Kuc2011}, as shown in Fig. \ref{fig:2}.
Such situation can be achieved by electro-gating \cite{Ye2012,Taniguchi2012}.
The effective Hamiltonian $\mathcal{H}_{0,\textrm{v}}$ for electrons in the $\textrm{v}K$ valley is given by \cite{Wang2012,Yuan2013}
\begin{align}
	\label{eq:2-2} 
& \mathcal{H}_{0,\textrm{v}}
 = \sum_{\bm{k}} \psi^\dagger_{\textrm{v}} \lf[ \varepsilon_{k} \sigma^0  + \textrm{v} \beta_{\textrm{I}} \sigma^z 
 	+ \alpha_{\textrm{R}} (k_y\sigma^x-k_x\sigma^y) \rf]\psi_{\textrm{v}},
\end{align}
where 
$\psi^\dagger_{\textrm{v}} \equiv \psi^\dagger_{\textrm{v}} (\bm{k}) = (\psi^\dagger_{\uparrow,\textrm{v}} \  \psi^\dagger_{\downarrow,\textrm{v}})$ is the creation operator of an electron in the valley $\textrm{v}K$ with $\uparrow$ and $\downarrow$ denoting the spin,
$\bm k\equiv(k_x,k_y)$ is the electron momentum measured from the $\textrm{v}K$ point,
and $\sigma^i\ (i=x,y,z)$ are the Pauli matrices.
The first term of Eq. (\ref{eq:2-2}) is the kinetic term with $ \varepsilon_{k} = \hbar^2k^2 /(2m) - \mu$, 
where the chemical potential $\mu$ is measured from the averaged energy of the spin-split conduction bands at the $\textrm{v}K$ point.
The second term is the Ising SOC. 
The coupling strength $\beta_\textrm{I}$ is assumed to be a constant \cite{NYuan2014,Lu2015}, since the spin spliting is independent of $\bm{k}$ up to the second order near the $\pm K$ points \cite{Xiao2012,Kormanyos2013,Rostami2013,Yuan2013,Ochoa2013,Liu2013}.
The third term of Eq. (\ref{eq:2-2}) is the Rashba SOC whose strength $\alpha_\textrm{R}$ can be controlled by the gate voltage. Without loss of generality, we choose both $\beta_\textrm{I}$ and $\alpha_\textrm{R}$ to be positive. Importantly, we note that in the presence of both Rashba and Ising SOCs, the effective Hamiltonian in Eq. (\ref{eq:2-2}) has the form of a massive Dirac Hamiltonian (by ignoring the $\varepsilon_{k}$-term which does not affect the Berry curvature). As we show in later sections, this massive-Dirac-like Hamiltonian leads to non-trivial valley-dependent Berry curvatures, which plays an essential role in the VEEs.

We further take into account an in-plane dc electric field $\bm E$ as well as nonmagnetic impurities on the MTMD. 
Here, we assume that hybridization between the $\pm K$ valleys can be ignored, namely, 
the magnitude of the momentum-shift due to $\bm{E}$ is much smaller than $2|K|$.
Then the Hamiltonian for electrons in the $\pm K$ valleys are decoupled, each of which is given by
\eq{
\label{eq:2-3} 
\mathcal{H}_{\textrm{v}}  = & \mathcal{H}_{0,\textrm{v}} + \mathcal{H}_{\textrm{em},\textrm{v}} + \mathcal{V}_{\textrm{imp},\textrm{v}},
	\\
\label{eq:2-4} 
\mathcal{H}_{\textrm{em},\textrm{v}} = & - e \sum_{\bm{k}} \psi^\dagger_{\textrm{v}} \bm{v}  \cdot \bm{A} \psi_{\textrm{v}},	
	\\
\label{eq:2-5} 
\mathcal{V}_{\textrm{imp},\textrm{v}} 
	  =& \int d\bm{x} u_\textrm{i} (\bm{x}) \psi^\dagger_{\textrm{v}}(\bm{x}) \psi_{\textrm{v}} (\bm{x}),
}
where $\bm{v} = - (\partial \mathcal{H}_{\textrm{v}}/\partial \bm{k})/\hbar $ is the velocity operator, $\bm A$ is the vector potential defined by $\bm{E}=-\partial_t \bm{A}$, 
and $u_{\textrm{i}}(\bm{x})=\sum_{j=1}^{N_{\textrm{i}}} u_0 \delta(\bm{x}-{\bm{R}_j }) $ is the short-range impurity  potential independent of the valley index.   
Here, $N_{\textrm{i}}$, $u_0$, and $\bm{R}_j$ are the number of impurities, a constant impurity potential, and the position of the $j$th impurity on the monolayer, respectively.

\begin{figure}[t] 
\centering
\includegraphics[width=7.cm]{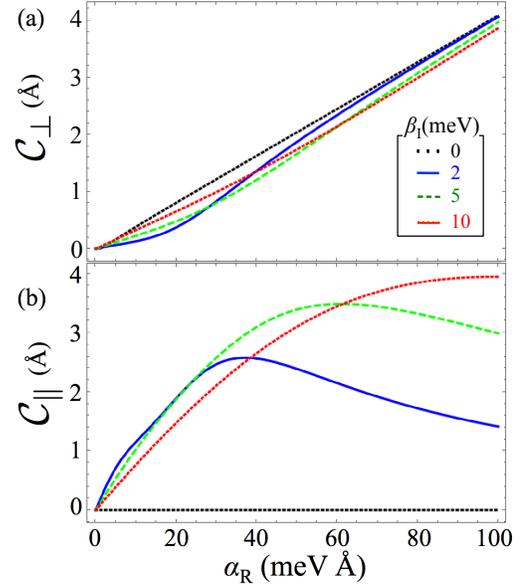}
\caption{ (Color online) 
(a) Conventional and (b) valley- dependent part of the VEE as a function of the Rashba SOC ($\alpha_{\textrm{R}}$) for various values of the Ising SOC ($\beta_{\textrm{I}}$) at bare self- energy $\Sigma_0 = 3$ meV and chemical potential $\mu$ = 100 meV.
}
\label{fig:3} 
\end{figure}

\textbf{Valley-Edelstein effect.}--
We calculate the induced spin density using the Keldysh Green's functions within the linear response to $\bm{E}$.
The contributions from the $\pm K$ valleys, $\langle \bm{s}^\textrm{VEE}_{\pm}\rangle$, are independently calculated from $\mathcal{H}_\pm$. The calculated perpendicular ($\mathcal{C}_{\perp}$) and parallel ($\mathcal{C}_{\parallel}$) spin coefficients are shown in Fig. \ref{fig:3}. Clearly, the unconventional $\mathcal{C}_{\parallel}$-term arises when both Rashba and Ising SOCs are present. According to Eq. (\ref{eq:1-2}), the parallel spin polarization for electrons from opposite valleys points to opposite directions. This is referred to as the valley Edelstein effect (VEE).

In our calculations, we assume that the self-energy $\Sigma_\textrm{v}$ due to impurity scatterings satisfies 
$\Sigma_\textrm{v} \ll |\mu\pm \beta_\textrm{I}|$, which allows us to take into account disorder effects perturbatively \cite{Edelstein1990}.
For the parameters we choose below, this assumption is satisfied. 
Using the Keldysh techniques, the spin density in response to $\bm E$ is found to be \cite{Haug2008}
\begin{align}
\label{eq:3-5} 
\langle s_{\textrm{v}}^{\textrm{VEE}, i} \rangle  
	= -  \frac{e\hbar}{4\pi}\sum_{\bm{k},\omega} \sum_{j=x,y} \frac{df_\omega}{d\omega}  \textrm{tr} 
				\bigl[
				\sigma^i
				G^{\rm{r}}_{\bm{k},\omega,\textrm{v}}
				\mathcal{S}^j_{\bm{k},\omega,\textrm{v}}
				G^{\rm{a}}_{\bm{k},\omega,\textrm{v}}
				\bigr] E_{j},
\end{align}
where $G^{\rm r}_{\bm{k},\omega,\textrm{v}}= [\hbar \omega - \mathcal{H}_{0,\textrm{v}} + i \Sigma_{\textrm{v}} ]^{-1}$ and $G^{\rm a}_{\bm{k},\omega,\textrm{v}}=[G^{\rm r}_{\bm{k},\omega,\textrm{v}}]^\dagger$ are the retarded and advanced Green's functions, respectively, $f_\omega$ is the Fermi distribution function, and $\mathcal{S}^j_{\bm{k},\omega,\textrm{v}}$ is the velocity operator with the ladder vertex corrections. 
The self-energy $\Sigma_{\textrm{v}}$ is calculated within the self-consistent Born approximation, resulting in \cite{Haug2008}   
\begin{align}
\label{eq:self-energy}
\Sigma_{\textrm{v}} & = \Sigma'_{0,\omega} +\textrm{v}  \Sigma_{z,\omega} \sigma^z, 
\end{align}
where  
$\Sigma'_{0,\omega}=\Sigma_0 \equiv  \pi n_{\rm{c}}u_{0}^2 \nu_e$ and $\Sigma_{z,\omega}=0$ for $\mu>\beta_\textrm{I}$ 
and 
$\Sigma'_{0,\omega} \equiv  \Sigma_0 (1 + u_{\textrm{R}}/ \sqrt{\lambda_\omega} )/2 $ 
and 
$\Sigma_{z,\omega} \equiv \Sigma_0  \beta_{\textrm{I}}/( 2\sqrt{\lambda_\omega}) $ 
for $-\beta_\textrm{I}<\mu<\beta_\textrm{I}$. 
Here, we define $\Sigma_0=  \pi n_{\rm{c}}u_{0}^2 \nu_e$ and 
$\lambda_\omega= \beta_{\textrm{I}}^2 -(\mu+\hbar\omega)^2 + [u_{\textrm{R}} - (\mu+\hbar\omega)]^2 $, where 
$n_{\rm{c}}$ is the concentration of impurities and 
$u_{\textrm{R}}\equiv m\alpha_{\textrm{R}}^2/\hbar^2$ is the Rashba energy. 

Using the obtained self-energy, the Green's function can be decomposed as
\begin{align}
	\nt 
G^{\rm r}_{\bm{k},\omega,\textrm{v}}
 =& \frac{\Omega^{\rm r}_+}{\hbar \omega - E_+ + i (\Sigma'_{0,\omega} +\gamma \Sigma_{z,\omega} ) } \\ \label{eq:GF}
  & + \frac{\Omega^{\rm r}_-}{\hbar \omega - E_- + i (\Sigma'_{0,\omega} - \gamma \Sigma_{z,\omega} ) },
\end{align}
where $E_\pm =\varepsilon_{k} \pm \sqrt{\alpha^2_{\textrm{R}}k^2 + \beta_{\textrm{I}}^2 }$ is the energy dispersion of each band, 
and 
$\Omega_\pm^{\textrm{r}} = 1/2 \pm (\bm{u}^{\textrm{r}} \cdot \bm{\sigma} )/ 2( \sqrt{\alpha^2_{\textrm{R}} k^2 + \beta_{\textrm{I}}^2} -i\gamma \Sigma_{z,\omega})$ 
is the projection operator onto the each band with $\bm{u}^{\textrm{r}}=\alpha_{\textrm{R}} (\bm{k}\times \hat{\bm z}) + \textrm{v}(\beta_{\textrm{I}}- i  \Sigma_{z}) \hat{\bm z}$ and $\gamma= \beta_{\textrm{I}}/\sqrt{\beta_{\textrm{I}}^2 + \alpha_{\textrm{R}}^2 k^2 }$.
It is noticed that the first (second) term of Eq. (\ref{eq:GF}) is the Green's function corresponding to the spin-split upper (lower) conduction band. 

After some calculations \cite{SupplementaryTMD}, 
we find that the magnitude of $\mathcal{C}_{\perp}$ is comparable to that of $\mathcal{C}_{\parallel} $ in Eq. (\ref{eq:1-2}) with $\mu \gg \beta_{\textrm{I}}$. Below, we discuss in detail in this parameter regime. 
The results for $\mu<\beta_{\textrm{I}}$ will be given in the Supplementary \cite{SupplementaryTMD}.
For $\mu\gg\beta_{\textrm{I}}$, we obtain
\eq{
\label{eq:coefficient-perp1}
& \mathcal{C}_{\perp} 
	 = \frac{\alpha_{\textrm{R}}}{4\pi} \Gamma^{(v)} \frac{1-\Gamma^{(s)}_{xx}}{[1 - \Gamma^{(s)}_{xx} ]^2 +  [\Gamma^{(s)}_{xy}]^2 },
	\\
\label{eq:coefficient-parallel1}
& \mathcal{C}_{\parallel} 
	 = \frac{\alpha_{\textrm{R}}}{4\pi} \Gamma^{(v)} \frac{ \Gamma^{(s)}_{xy}}{[1 - \Gamma^{(s)}_{xx} ]^2 +  [\Gamma^{(s)}_{xy}]^2 }
}
where $\Gamma^{(s)}_{xx}$ ($\Gamma^{(s)}_{xy}$) is the diagonal (off-diagonal) component of the spin-vertex function, and $\Gamma^{(v)}$ is the velocity-vertex function, which are defined from the following equations: 
$n_{\rm{c}}u_{\textrm{i}}^2 \sum_{\bm{k}} G^{\rm{r}}_{\bm{k},0,\textrm{v}} \sigma^x G^{\rm{a}}_{\bm{k},0,\textrm{v}} 
  =\textrm{v} \Gamma_{xx}^{(\textcolor{red}{s})} \sigma^{x} + \Gamma_{xy}^{(s)} \sigma^{y}$ 
and 
$  \sum_{\bm{k}}  G^{\rm{r}}_{\bm{k},0,\textrm{v}} v^j G^{\rm{a}}_{\bm{k},0,\textrm{v}} 
= \Gamma^{(v)} \epsilon_{j\ell z} \sigma^\ell  \alpha_{\textrm{R}}\nu_e /(2\hbar)$.

We numerically calculate the vertex functions ($\Gamma^{(v)}$, $\Gamma^{(s)}_{xx}$, and $\Gamma^{(s)}_{xy}$) 
and obtain the Rashba SOC ($\alpha_\textrm{R}$) and Ising SOC ($\beta_\textrm{I}$) dependence of $\mathcal{C}_\perp$ and $\mathcal{C}_\parallel$ as shown in Fig. \ref{fig:3}(a) and \ref{fig:3}(b), respectively.
Here, we choose $\mu=100$ meV and $\Sigma_0=3$ meV (which corresponds to the relaxation time $\tau=0.1$ ps).

The results presented in Fig. \ref{fig:3} exhibit two important features. First, at $\beta_{\rm I}=0$, 
we find that $\mathcal{C}_\perp \approx \mathcal{C}_\perp^0 \equiv \alpha_{\textrm{R}}/(2\Sigma_0)$, and $\mathcal{C}_\parallel = 0$. Namely, the conventional Edelstein effect generated by Rashba SOCs is reproduced. Second, in $\beta_{\textrm{I}} \neq 0$, the $\mathcal{C}_\perp$-term is suppressed, while the $\mathcal{C}_\parallel$-term becomes non-zero. For fixed $\beta_{\textrm{I}}$, $\mathcal{C}_{\parallel}$ increases as a function of $\alpha_{\textrm{R}}$ and reaches a maximum value where $\mathcal{C}_\parallel \sim \mathcal{C}_\perp$. With further increase in $\alpha_{\textrm{R}}$, $\mathcal{C}_\parallel$ decreases and $\mathcal{C}_\perp$ dominates the Edelstein effect.

To understand this unusual behavior of $\mathcal{C}_{\parallel}$, we note that it can be approximated as $\mathcal{C}_{\parallel}  \approx \frac{\alpha_{\textrm{R}} }{4\pi } \Gamma^{(v)} \Gamma^{(s)}_{xy}$, where $\Gamma_{xy}^{(s)} \approx  \frac{ 2 \beta_{\textrm{I}} \Sigma_0}  { \alpha_{\textrm{R}}^2 k_F^2+\beta_{\textrm{I}}^2  }$ 
 and 
 $\Gamma^{(v)} \approx \frac{2\pi}{\Sigma_0} \lf( 	1-  \frac{1}{2}\frac{\alpha^2_{\textrm{R}} k_F^2 }{ \alpha^2_{\textrm{R}} k_F^2 +\beta_{\textrm{I}}^2  } \rf)$. Here, $k_F$ refers to the Fermi momentum measured from the $K$-points.
Remarkably, the expression of $\mathcal{C}_{\parallel}$ can be recast in the following form:
\begin{eqnarray}\label{eq:C_parallel}
\mathcal{C}_{\parallel} \simeq  \cos(\theta_{\parallel})|\Omega_{\textrm{spin}}^{\textrm{v}=\pm}(k_{\textrm{F}})|k_{\textrm{F}} \lf(1 + \frac{2\beta^2_{\textrm{I}}}{\alpha^2_{\textrm{R}}k^2_\textrm{F}}\rf).
\end{eqnarray}

\begin{figure}[tb]
\includegraphics[width=8.7cm]{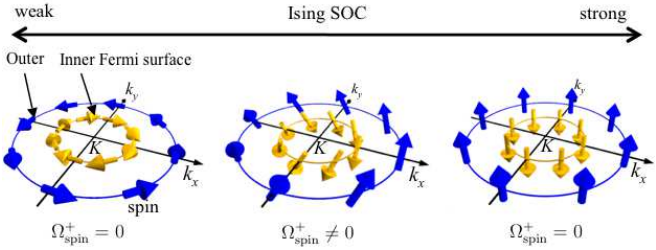}
\caption{ (Color online)
Schematic of the spin structures at the Fermi surface. The Berry curvature $\Omega^{\textrm{v}=+}_{\textrm{spin}}$ induced by Rashba and Ising SOCs can be visualized from the solid angle of the spin structures at $K$-point: The Berry curvature is nearly zero at weak (left panel) and strong (right panel) Ising SOC. Finite Berry curvature can emerge when Ising and Rashba SOCs are comparable. 
}
\label{fig:4} 
\end{figure}

Here, $\cos(\theta_{\parallel}) = \alpha_{\textrm{R}}k_\textrm{F}/( \alpha_{\textrm{R}}^2 k_{\textrm{F}}^2  + \beta_{\textrm{I}}^2 )^{1/2}$ is the in-plane direction cosine of electron spin at the Fermi energy. $\Omega_{\textrm{spin}}^{\textrm{v}=\pm}$ is the Berry curvature based on the massive-Dirac-like Hamiltonian in Eq. (\ref{eq:2-2}): 
\eq{\label{eq:BerryCurvature}
\Omega^{\textrm{v}=\pm}_{\textrm{spin}} (k_\textrm{F}) = \textrm{v} \frac{\alpha^2_{\textrm{R}} \beta_{\textrm{I}} }{2( \alpha_{\textrm{R}}^2 k_{\textrm{F}}^2  + \beta_{\textrm{I}}^2 )^{3/2}} .
} 
Interestingly, the valley-index in $\Omega^{\textrm{v}}_{\textrm{spin}}$ results from the Ising SOC, which plays the role of a valley-dependent Dirac mass in Eq. (\ref{eq:2-2}). Based on Eq. (\ref{eq:BerryCurvature}), the magnitude of $\Omega^{\textrm{v}}_{\textrm{spin}}$ is a non-monotonic function of $\alpha_{\textrm{R}}$ and $\beta_{\textrm{I}}$, which can be visualized from the solid angle of the spin structures at the Fermi surface (Fig. \ref{fig:4}): When either $\beta_{\textrm{I}}$ or $\alpha_{\textrm{R}}$ is zero, 
the spin structure is either coplanar (left panel) or uniformly out-of-plane (right panel). 
In either case, $\Omega^{\textrm{v}}_{\textrm{spin}}$ is zero. 
In contrast, for $\alpha_{\textrm{R}}k_{\textrm{F}}\sim\beta_{\textrm{I}}$ (middle panel), $\Omega^{\textrm{v}}_{\textrm{spin}}$ is nonzero, which results in a finite $\mathcal{C}_{\parallel}$. Notably, this special behavior of $\Omega^{\textrm{v}}_{\textrm{spin}}$ is qualitatively consistent with the non-monotonic behavior of $\mathcal{C}_{\parallel}$ as a function of $\alpha_\textrm{R}$ in Fig. \ref{fig:3}: when either $\alpha_{\textrm{R}}k_{\textrm{F}}\gg\beta_{\textrm{I}}$ or $\alpha_{\textrm{R}}k_{\textrm{F}}\ll\beta_{\textrm{I}}$, $\Omega^{\textrm{v}}_{\textrm{spin}} \approx 0$ and $\mathcal{C}_{\parallel}$ is small. In the intermediate regime, $\Omega^{\textrm{v}}_{\textrm{spin}}$ can be strong enough to induce a large $\mathcal{C}_{\parallel}$, with $\mathcal{C}_{\parallel}\sim \mathcal{C}_{\perp}$. Based on the parameters in Fig. \ref{fig:3}, the Berry curvature in Eq. (\ref{eq:BerryCurvature}) is estimated to be $\Omega^{\textrm{v}}_{\textrm{spin}} \approx 1 $ \AA$^2$ at the Fermi energy with $\mu = 100 \ \textrm{meV}, \alpha_{\textrm{R}}=20 \ \textrm{meV}$ \AA, which is ten times of the intrinsic Berry curvature $\Omega_{\textrm{orbital}} \approx 0.1 $ \AA$^2$ with the same Fermi momentum $k_{\textrm{F}}$ \cite{SupplementaryTMD}.

The close relation between $\mathcal{C}_{\perp}$ and the Berry curvature in Eq. (\ref{eq:BerryCurvature}) reveals the physical origin of VEEs as a combined effect of the valley-dependent $\Omega^{\textrm{v}}_{\textrm{spin}}$ and conventional Edelstein effect: Under applied electric fields, $\Omega^{\textrm{v}}_{\textrm{spin}}$ drives electrons from opposite valleys to drift in opposite transverse $\textit{y}$-directions (Fig. \ref{Fig1}(a)). The resultant transverse flow of electrons from opposite valleys combine with conventional Edelstein effects to induce valley-contrasting spin polarizations that are parallel to the applied electric field.
\begin{figure}[b]
\centering
\includegraphics[width=8.6cm]{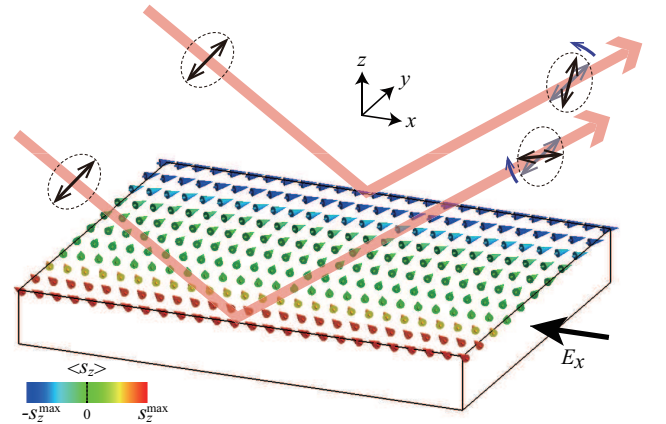}
\caption{\label{Fig5} (Color online) Schematic for spatial spin textures generated by VEEs and detection scheme using Kerr effects. Magnitude of out-of-plane spin component is qualitatively indicated by colors. Parallel spins on the edges can be detected by the Kerr angle $\theta^{\textrm{s}}_{\textrm{K}}$ [Eq. (\ref{eq:Kerr})].
}
\end{figure}
%

\textbf{Experimental realization and detection.}--
In this section, we discuss how to experimentally realize and detect VEEs in gated MTMDs. Particularly, we propose that the parallel spin induced by VEEs can be detected by longitudinal Kerr effect measurements \cite{You1998}.

Consider the MTMD system in Fig. \ref{Fig1}(b), by applying an electric field in the \textit{x}-direction, electrons from opposite valleys are driven by $\Omega^{\textrm{v}}_{\textrm{spin}}$ to drift in opposite $y$-directions. In the steady state, this establishes a valley imbalance near the boundaries \cite{Xiao2012}, where finite spin density due to VEEs will also emerge (Fig. \ref{Fig5}).

Here, we note that VEE has two unique signatures. First, the induced magnetization $\bm{M}_{\parallel}$ is parallel to $\bm{E}$, which contrasts with the in-plane perpendicular magnetization from conventional Edelstein effect. Second, the non-zero $\bm{M}_{\parallel}$ induced by VEEs are valley-dependent (Eq. (\ref{eq:1-2})). Due to the valley Hall effect resulting from $\Omega^{\textrm{v}}_{\textrm{spin}}$, valley polarizations accumulated near opposite edges have opposite signs \cite{Xiao2012}. As a result, $\bm{M}_{\parallel}$ due to VEEs also point to opposite directions at opposite edges. Therefore, observation of non-zero edge-contrasting $\bm{M}_{\parallel}$ provides strong evidence for VEEs.

Now, we discuss how to detect $\bm{M}_{\parallel}$ using longitudinal Kerr effect measurements. In magneto-optic Kerr effect, an incident light of $\textrm{s}(\textrm{p})$-polarizations are generally reflected as superposition of $\textrm{s}$- and $\textrm{p}$-polarized lights due to a magnetized surface. This effect is quantified by the Kerr angles $\theta^{i}_{\textrm{K}} \ (i=\textrm{s},\textrm{p})$ for $i$-polarized lights \cite{You1998}. It can be shown that with proper oblique incidence setting, $\bm{M}_{\parallel}$ can be related to the Kerr angle for $\textrm{s}$-polarized light \cite{You1998, SupplementaryTMD}:
\begin{eqnarray}\label{eq:Kerr}
\hat{x} \cdot \bm{M}_{\parallel}/M_{\textrm{tot}} \propto \theta^{\textrm{s}}_{\textrm{K}}.
\end{eqnarray}
Here, $M_{\textrm{tot}}$ is the magnitude of the total magnetization. Moreover, the edge-contrasting $\bm{M}_{\parallel}$ from VEEs can be mapped out by the spatial profile of $\theta^{\textrm{s}}_{\textrm{K}}$, where the valley-dependent spin density can be signified by opposite signs of $\theta^{\textrm{s}}_{\textrm{K}}$ at opposite edges. Details of Kerr effect setting can be found in the Supplementary Materials \cite{SupplementaryTMD}.

\textbf{Conclusion.}--In this work, we predict that Berry curvatures due to coexisting Rashba and Ising SOCs combined with conventional Edelstein effects lead to VEEs in gated MTMDs, in which valley-contrasting spin polarization parallel to the applied electric field can be generated. The parallel spin polarization due to VEEs can be comparable to the perpendicular spin polarization via conventional Edelstein effect. Experimental realization of VEE can be detected by longitudinal Kerr effects.

\begin*{acknowledgments}
We acknowledge useful discussions with T. Yokoyama, T. Takenobu, C.-Z. Chen and K. F. Mak.
This work was supported by a Grant-in-Aid for the Core Research for Evolutional Science and Technology (CREST) of the Japan Science and Technology Corporation (JST) [JPMJCR14F1] and by JSPS KAKENHI Grant Numbers JP15K17726, JP16H00989; 
K.T. is supported by a Grant-in-Aid for JSPS Fellows (Grants No. 13J03141). K. T. L thanks the support of HKRGC, the Croucher Foundation and Dr. Tai-chin Lo Foundation through HKUST3/CRF/13G, C6026-16W, 16324216 and Croucher Innovation Grants.

\end*{acknowledgments}

\newpage

\begin{widetext}
\begin{center}
\textbf{\large 
Supplemental Material for "Valley Edelstein Effect in Monolayer Transition Metal Dichalcogenides"}
\end{center}

\begin{center}
Katsuhisa Taguchi$^{1}$, Benjamin T. Zhou$^{2}$, Yuki Kawaguchi$^{1}$, Yukio Tanaka$^{1}$, and K. T. Law$^{2}$

\it{$^1$Department of Applied Physics, Nagoya University, Nagoya, 464-8603, Japan}

\it{$^2$Department of Physics, Hong Kong University of Science and Technology, Clear Water Bay, Hong Kong, China
}
\end{center}
\setcounter{equation}{0}
\setcounter{figure}{0}
\setcounter{table}{0}
\setcounter{page}{1}
\makeatletter
\renewcommand{\theequation}{S\arabic{equation}}
\renewcommand{\figurename}{Figure S}

\
\\
In this supplementary material, 
we first present the magnitude of Berry curvature in gated (monolayer transition metal dichalcogenides) MTMDs.
Second, we provide the calculation of $\mathcal{C}_{\parallel}$ and $\mathcal{C}_{\perp}$ for $\mu > \beta_{\textrm{I}} $. 
Third, we show the details of the calculation of $\mathcal{C}_{\parallel}$ and $\mathcal{C}_{\perp}$ for $\mu > \beta_{\textrm{I}} $.
In addition, we present the relation between $\mathcal{C}_{\parallel}$ and the Berry curvature in gated MTMDs.  
Here, $\mathcal{C}_{\parallel}$ and $\mathcal{C}_{\perp}$ for $\mu > \beta_{\textrm{I}}  $ are represented in Eq. (\ref{eq:1-2}) in the main text. 
In section \ref{sec:detection_way}, we present details on the detection scheme of the valley Edelstein effect (VEE) by using Kerr effect measurements.
Finally, we estimate the induced spin density.

\section{Berry curvature in gated MTMDs}\label{sec:Berry curvature}
We evaluate Berry curvature in pristine MTDMs and that in gated MTMDs. 
The former is given by the effective massive Dirac-Hamiltonian. 
The effective Hamiltonian is given by $\mathcal{H}_{\textrm{orbital}}= V_F(\textrm{v} k_x \tau_x + k_y \tau_y) + \Delta \tau_z$, where $\textrm{v}=\pm$ is the valley index, $V_F$ is the Fermi velocity, $\tau$ is Pauli matrix acting on orbital degrees of freedom, and $2\Delta$ corresponds to the band gap between conduction band and valence band \cite{Xiao2012}.
Then, the intrinsic Berry curvature $\Omega_{\textrm{orbital}}$ is given by 
\eq{
\Omega_{\textrm{orbital}} (k) = \frac{V_F^2\Delta}{2(V_F^2 k^2 + \Delta^2)^{3/2}}. 
}
On the other hand, the Berry curvature in gated MTMDs results from the effective massive-Dirac-like Hamiltonian given by 
$\mathcal{H}_\textrm{spin}= (\frac{|k|^2}{2m}-\mu)\sigma_0+ \alpha_{\textrm{R}}(k_y \sigma_x - k_x \sigma_y) + \textrm{v}\beta_{\textrm{I}}\sigma_z$ which is also a Dirac-type Hamiltonian, but with an extra $\sigma_0$-term which does not affect the Berry curvature. Here, $\sigma$ is the Pauli matrix in spin space. 
The Berry curvature induced by Rashba and Ising SOCs is given by 
\eq{
\Omega^{\textrm{v}=\pm}_{\textrm{spin}} (k)
	& \equiv \hat{z}\cdot\bm{\nabla} \times \langle \bm{k},\textrm{v} | i\bm{\nabla} |\bm{k}, \textrm{v} \rangle
	= \textrm{v} \frac{\alpha_{\textrm{R}}^2 \beta_{\textrm{I}} }{2( \alpha_{\textrm{R}}^2 k_{\textrm{F}}^2  + \beta_{\textrm{I}}^2 )^{3/2}}, 
} 
where $|\bm{k}, \textrm{v} \rangle$ is the wave function of the effective massive Dirac-Hamiltonian in the gated TMD. 
We find that the Berry curvature in gated MTMD is estimated by 
$|\Omega^{\textrm{v}=\pm}_{\textrm{spin}} (k=k_{\textrm{F}})| \approx 1.26$ \AA$^2$, when we use realistic parameter, 
$\beta_{\textrm{I}}=10$ meV, $\alpha_{\textrm{R}}=20$ meV \AA, $\mu=100$ meV, and $m/m_e = 0.5$ with $m_e$ being the electron rest mass. 
On the other hand, the intrinsic Berry curvature is given by $|\Omega^{\textrm{v}=\pm}_{\textrm{orbital}} (k=k_{\textrm{F}})| \approx 0.096$ \AA$^2$, when we chose the realistic parameter, $\Delta=1.79$ eV, $V_F=4.38$ eV \AA, $\mu=100$ meV, and $m/m_e = 0.5$. Therefore, we find that in the regime considered in this work, 
$|\Omega^{\textrm{v}=\pm}_{\textrm{spin}} (k_{\textrm{F}})| \gg |\Omega_{\textrm{orbital}} (k_{\textrm{F}})|$.

\begin{figure}[b]
\includegraphics[width=8.5cm]{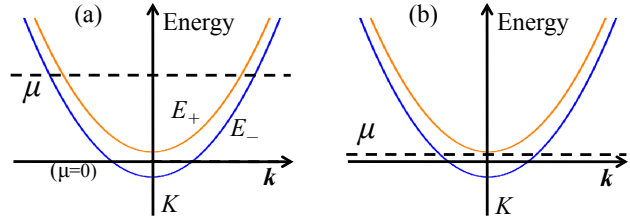}
\caption{ Schematic illustration of spin-split conduction bands of the MTMD around $K$-point  
 for (a) $\mu  > \beta_{\textrm{I}}$ and (b) $ - \beta_{\textrm{I}}  \leq \mu  \leq \beta_{\textrm{I}} $. 
}
\label{figS2} 
\end{figure} 

\section{Derivation of $\mathcal{C}_{\parallel}$ and  $\mathcal{C}_{\perp}$ in $|\mu - \Sigma_0| < \beta_{\textrm{I}} $ }\label{sec:I-REE}
Since preexisting works of the conventional Edelstein effect used Green's function techniques\cite{Edelstein1990,Inoue2002}, we also use Green's functions in the following calculation. 
The calculation is assumed when the magnitude of the self-energy of nonmagnetic impurity scattering is smaller than that of the chemical potential ($\mu \gg \Sigma_0$).

First, we introduce the impurity-averaged Green's functions in $|\mu - \Sigma_0| < \beta_{\textrm{I}} $: 
\eq{ \label{eq:GF1R}
G^{\textrm{r}}_{\bm{k},\omega,\textrm{v}}
	& = \frac{ \Omega_{+}^\textrm{r} }{ \hbar\omega - E_+ +  i ( \Sigma'_0 + \gamma \Sigma_{z} ) }
	+
	\frac{ \Omega_{-}^\textrm{r} }{ \hbar\omega - E_- +  i ( \Sigma'_0 - \gamma \Sigma_{z} ) }, 
	\\
	\label{eq:GF1A}
G^{\textrm{a}}_{\bm{k},\omega, \textrm{v}}
	& = \frac{ \Omega_{+}^\textrm{a} }{ \hbar\omega - E_+ -  i ( \Sigma'_0 + \gamma \Sigma_{z} ) }
	+
	\frac{ \Omega_{-}^\textrm{a} }{ \hbar\omega - E_- - i ( \Sigma'_0 - \gamma \Sigma_{z} ) },
}
with 
\eq{
&\Omega_\pm^{\textrm{r}} 	 = \frac{1}{2} \lf[ 1 \pm \frac{ \bm{u}^{\textrm{r}}\cdot\bm{\sigma} }{u_0^{\textrm{r}}}\rf],& 
&u_0^{\textrm{r}}  \equiv \sqrt{\alpha^2_{\textrm{R}} k^2 + \beta_{\textrm{I}}^2} -i\gamma \Sigma_z, &
	\\
&\Omega_\pm^{\textrm{a}} = \frac{1}{2} \lf[ 1 \pm \frac{\bm{u}^{\textrm{a}}\cdot\bm{\sigma} }{u_0^{\textrm{a}}}\rf],& 
&u_0^{\textrm{a}}  \equiv \sqrt{\alpha^2_{\textrm{R}} k^2 + \beta_{\textrm{I}}^2} +i\gamma \Sigma_z. &
}
Here, we define 
$\gamma= \beta_{\textrm{I}}/\sqrt{ \alpha_{\textrm{R}}^2 k^2 + \beta_{\textrm{I}}^2  }$
and 
$\bm{u}^{\textrm{r}}=\alpha_{\textrm{R}} (\bm{k}\times \hat{\bm z}) + \textrm{v}(\beta_{\textrm{I}}- i  \Sigma_{z}) \hat{\bm z}$, which satisfies $\bm{u}^{\textrm{r}} [= (\bm{u}^{\textrm{a}})^*]$.
Here, $E_{\pm} $ denotes the energy dispersion of the spin-splitting bands 
$E_{\pm}  = \epsilon_{k}  \pm \sqrt{\alpha^2_{\textrm{R}} k^2 + \beta_{\textrm{I}}^2}$. 
It is noticed that the first (second) term of Eqs. (\ref{eq:GF1R})-(\ref{eq:GF1A}) is the Green's function corresponding to the upper (lower) spin split conduction band.
Since the top conduction band is far from the Fermi level, 
contributions from the first term of Eqs. (\ref{eq:GF1R})-(\ref{eq:GF1A}) in Eq. (1) in the main text are negligibly small compared with that from the second term of Eqs. (\ref{eq:GF1R})-(\ref{eq:GF1A}). Hence the first term of Eqs. (\ref{eq:GF1R})-(\ref{eq:GF1A}) is ignored in the following calculation.  
Besides, below, we simply use the following representation $\Omega_{-} \equiv \Omega_{-}^{\textrm{r}}$, $\Omega_{-}^\dagger \equiv \Omega_{-}^{\textrm{a}}$, $\bm{u} \equiv \bm{u}^{\textrm{r}}$, and $\bm{u}^* \equiv \bm{u}^{\textrm{a}}$. 

\begin{table}[t]
\caption{
\label{tab1}
Self-energy due to the impurity scatterings in the presence of Rashba SOC and Ising SOC 
where $\Sigma'_{0,\omega=0}$ ($\Sigma_{z,\omega=0}$) is the spin and valley independent (dependent) part of the self-energy at $\omega=0$, $\Sigma_0$ is the self-energy for a 2D metal, 
$u_{\rm{R}} \equiv m\alpha_{\rm{R}}^2/\hbar^2$, and $\lambda_0=\beta_{\rm{I}}^2+u_{\rm{R}}^2-2 u_{\rm{R}} \mu$. 
The dependence on the SOC strengths completely changes according to
whether the Fermi level crosses one ($ - \beta_{\textrm{I}}  \leq \mu  \leq \beta_{\textrm{I}} $) or two ($ \beta_{\textrm{I}} < \mu $) conduction bands. }
\begin{ruledtabular}
\begin{tabular}{ccc}
Fermi level  & $\Sigma'_{0,\omega=0}/\Sigma_{0}$ & $\Sigma_{z,\omega=0}/\Sigma_{0}$ \\
\hline
 $\mu > \beta_{\textrm{I}}  $  & $1$    & $0$ \\
 $ - \beta_{\textrm{I}}  \leq \mu  \leq \beta_{\textrm{I}}  $   & $\frac{1}{2} + \frac{u_{\textrm{R}}}{2\sqrt{\lambda_0}}  $    & $\frac{\beta_{\textrm{I}}}{2\sqrt{\lambda_0}}$  \\
\end{tabular}
\end{ruledtabular}
\end{table}

From Eq. (6) in the main text, 
the spin density of each valley due to the applied electric field is given by
\eq{ \label{eq:spin2-1}
\langle s^{\textrm{VEE},i}_{\textrm{v}} \rangle 
	 = -\frac{e\hbar}{4\pi} \sum_{\bm{k}} \sum_{j=x,y}
		\textrm{tr} \bigl[  \sigma^i G^{\textrm{r}}_{\bm{k}, \textrm{v}} \mathcal{S}^j_{\bm{k},0,\textrm{v}} G^{\textrm{a}}_{\bm{k}, \textrm{v}}\bigr] E_{j}. 
}
$ \mathcal{S}^j_{\bm{k},\omega,\textrm{v}} $ is defined  by 
\eq{
\mathcal{S}^j_{\bm{k},\omega,\textrm{v}}
	& =  v_j
	  + n_{\rm c}u_{\rm i}^2 \sum_{\bm{k}} 
	G^{\textrm{r}}_{\bm{k},\omega,\textrm{v}} v_j G^{\textrm{a}}_{\bm{k},\omega,\textrm{v}}
	  + 
	  (n_{\rm c}u_{\rm i}^2)^2  \sum_{\bm{k},\bm{k}_1}
G^{\textrm{r}}_{\bm{k}_1,\omega,\textrm{v}} G^{\textrm{r}}_{\bm{k},\omega,\textrm{v}} v_j G^{\textrm{a}}_{\bm{k},\omega,\textrm{v}} G^{\textrm{a}}_{\bm{k}_1,\omega,\textrm{v}}
	  +\cdots, 
}
where $v_j \equiv \partial \mathcal{H}_\textrm{v}/(\partial \hbar k_j) $ is the velocity operator. 
Here, $G^{\textrm{r}}_{\bm{k},v} [= (G^{\textrm{a}}_{\bm{k},v})^\dagger ]$ is defined by 
\eq{ 
G^{\textrm{r}}_{\bm{k},v} 
	& = \frac{ \Omega_{-} }{ - E_- + i (\Sigma'_0 - \gamma \Sigma_z)}	.
}
Below, we use the following $2\times2$ matrix
\eq{
	\label{eq:Spin-vertex-F}
& \Gamma_j^{(s)} \equiv n_{\rm c}u_{\rm i}^2 \sum_{\bm{k}} G^{\textrm{r}}_{\bm{k},\textrm{v}} \sigma^j G^{\textrm{a}}_{\bm{k},\textrm{v}}
	= \sum_{\ell=x,y} \Gamma^{(s)}_{n\ell} \sigma^\ell.
	\\
	\label{eq:Velocity-vertex-F}
&\tilde{\Gamma}_j^{(v)} \equiv \sum_{\bm{k}} G^{\textrm{r}}_{\bm{k},\textrm{v}} v_{j} G^{\textrm{a}}_{\bm{k},\textrm{v}}
	= \sum_{n=x,y} \Gamma^{(v)}_{jn} \sigma^n. 
}
Here, $\tilde{\Gamma}_j^{(s)} $ and $\Gamma_j^{(v)} $ denote the vertex function of the velocity operator and of the spin operator, respectively. 
$\tilde{\Gamma}_{jn}^{(s)}$ and $\Gamma_{jn}^{(v)}$ are coefficients of the matrix component of $\tilde{\Gamma}_j^{(s)}$ and $\Gamma_j^{(v)}$, respectively. 
Then,  
$\sum_{\bm{k}} G^{\textrm{r}}_{\bm{k},v} \mathcal{S}^j_{\bm{k},0,\textrm{v}} G^{\textrm{a}}_{\bm{k},\textrm{v}}$ 
of Eq. (\ref{eq:spin2-1}) is described by using $\Gamma_{jn}^{(v)}$ and $\Gamma_{jn}^{(s)}$ as 
\eq{ \nt 
\sum_{\bm{k}} G^{\textrm{r}}_{\bm{k},\textrm{v}} \mathcal{S}^j_{\bm{k},0,\textrm{v}} G^{\textrm{a}}_{\bm{k},\textrm{v}}
	& = \sum_{n=x,y} \tilde{\Gamma}^{(v)}_{jn}
		\biggl[ \sigma^n 
	+
	 n_{\rm c}u_{\rm i}^2 \sum_{\bm{k}} G^{\textrm{r}}_{\bm{k},\textrm{v}} \sigma^n G^{\textrm{a}}_{\bm{k},\textrm{v}}
	  + 
	  (n_{\rm c}u_{\rm i}^2)^2  \sum_{\bm{k},\bm{k}_1}
	  G^{\textrm{r}}_{\bm{k}_1,\textrm{v}} G^{\textrm{r}}_{\bm{k},\textrm{v}} \sigma^n G^{\textrm{a}}_{\bm{k},\textrm{v}}G^{\textrm{a}}_{\bm{k}_1,\textrm{v}}
	  +\cdots
		\biggr]
	\\ \label{eq:vertex-function1}
	& = \sum_{n,\ell=x,y} \tilde{\Gamma}^{(v)}_{jn}
		\lf[ \delta_{n\ell} + \Gamma^{(s)}_{n\ell} + \Gamma^{(s)}_{n\ell_1}\Gamma^{(s)}_{\ell_1 \ell} +\cdots
		\rf] \sigma^\ell. 
}
As a result, the electric field-induced spin density $\langle s^{\textrm{VEE},i}_{\textrm{v}} \rangle $ is given from Eqs. (\ref{eq:spin2-1})-(\ref{eq:vertex-function1}) as 
\eq{\nt 
\langle \bm{s}^{\textrm{VEE}}_{\textrm{v}} \rangle 
	& =  - \frac{e\hbar}{2\pi} \sum_{n,j=x,y} \tilde{\Gamma}^{(v)}_{jn} 
		\lf[ \delta_{n i} + \Gamma^{(s)}_{ni} + \Gamma^{(s)}_{n\ell_1}\Gamma^{(s)}_{\ell_1 i} +\cdots
		\rf] 	E_{j}. 
		\\ \label{eq:1-2-a} 
	& =- e \nu_e \mathcal{C}_\perp (\hat{\bm{z}}\times \bm{E}) 
	 	- \textrm{v} e \nu_e \mathcal{C}_\parallel \bm{E}, 
}
where 
$\mathcal{C}_{\perp} $ and $\mathcal{C}_{\parallel} $ are given by 
\eq{
\mathcal{C}_{\perp} 
	& = \frac{\alpha_{\textrm{R}}}{4\pi} \Gamma^{(v)}_{xy} [ 1- \Gamma^{(\textrm{s})} ]^{-1}, 
	\\ 
\mathcal{C}_{\parallel} 
	& = \frac{\alpha_{\textrm{R}}}{4\pi} \Gamma^{(v)}_{xx} [ 1- \Gamma^{(\textrm{s})} ]^{-1} 
}
with 
\eq{
\Gamma_{xy}^{(v)} & \equiv 
	\int_0^\infty   \frac{1}{ E_{\xi}^2 + (\Sigma'_0 - \gamma \Sigma_z)^2}
	      \frac{\xi \sqrt{ 2u_{\textrm{R}} \xi + \beta_{\textrm{I}}^2}}{2u_{\textrm{R}} \xi  + \beta_{\textrm{I}}^2 + \gamma^2 \Sigma_z^2 } d\xi
	- {\frac{2\pi}{\Sigma_0}}\Gamma^{(\textrm{s})}, 
	\\
\Gamma_{xx}^{(v)} & \equiv
	\int_0^\infty   \frac{\Sigma_z}{ E_{\xi}^2 + (\Sigma'_0 -  \gamma \Sigma_z)^2}
		\frac{\xi  - \sqrt{2u_{\textrm{R}} \xi + \beta_{\textrm{I}}^2} + \gamma\beta_{\textrm{I}}}
			{ 2u_{\textrm{R}} \xi   + \beta_{\textrm{I}}^2 + \gamma^2 \Sigma_z^2} d\xi, 
	\\ \nt 
\Gamma^{(\textrm{s})} 
	& = {\frac{\Sigma_0}{4\pi }} \int_0^\infty   
	 	\frac{1}{E_{\xi}^2 + (\Sigma'_0 - \gamma \Sigma_z)^2} 
		\lf[1 -  \frac{\beta_{\textrm{I}}^2 + \Sigma_z^2 }{2 u_{\textrm{R}} \xi  + \beta_{\textrm{I}}^2 + \gamma^2 \Sigma_z^2 }   \rf] d\xi, 
}
where we have used {$\Sigma_0 = \pi \nu_e n_{\textrm{c}} u_{\textrm{i}}^2 $}, 
$u_{\textrm{R}}=m\alpha^2_{\textrm{R}}/\hbar^2$, 
and 
$E_{\xi}  = \xi - \mu  + \sqrt{2 u_{\textrm{R}} \xi + \beta_{\textrm{I}}^2}$.
Here, $\Sigma'_0$, $\Sigma_{z}$, and $\gamma$ are given by 
\eq{ 
& \Sigma'_0  =  \biggl[ \frac{1}{2} + \frac{u_{\textrm{R}}}{2 \sqrt{\lambda_0}} \biggr] \Sigma_0, &
& \Sigma_z  =  \frac{\beta_{\textrm{I}}}{2\sqrt{\lambda_0}} \Sigma_0, & 
&\lambda_0 = \beta_{\textrm{I}}^2 - \mu^2 + (u_{\textrm{R}} - \mu)^2  ,&
& \gamma (\xi) = \frac{ \beta_{\textrm{I}} }{ \sqrt{2u_{\textrm{R}} \xi +\beta_{\textrm{I}}^2 }}.&
}

Figure S \ref{fig:S1} shows that the Ising SOC dependence of $\mathcal{C}_\parallel /\mathcal{C}_\perp $ in $\beta_{\textrm{I}} < \mu$ for several $\alpha_{\textrm{R}}$.
We find $\mathcal{C}_\parallel /\mathcal{C}_\perp <1$ in the whole of $\beta_{\textrm{I}}$. 
\begin{figure}[h]
  \includegraphics[width=9cm]{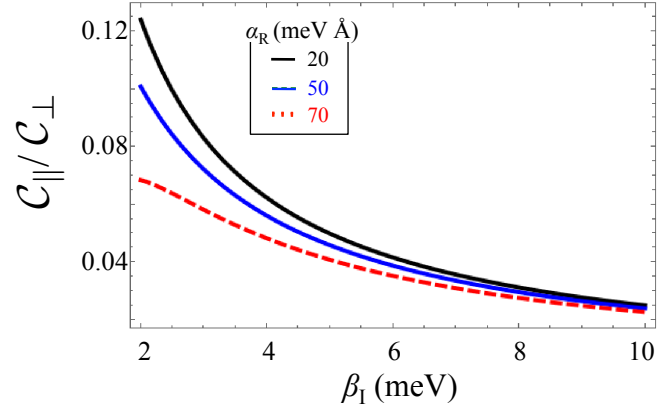}
  \caption{
The Ising SOC $\beta_\textrm{I}$ dependence of $\mathcal{C}_\parallel /\mathcal{C}_\perp $ in the conduction band in $\beta_\textrm{I} > \mu$ for several Rashba SOC ($\alpha_{\textrm{R}}$) at $\Sigma_0 =0.3$ meV.}
  \label{fig:S1} 
\end{figure} 

In addition, because of the spin-vertex function $|\Gamma_{xy}^{(s)}|<1$, 
$\mathcal{C}_{\parallel}$ and  $\mathcal{C}_{\perp}$ are approximately given by 
\eq{\nt
\mathcal{C}_{\perp} 
	& \approx \frac{\alpha_{\textrm{R}}}{4\pi} \Gamma^{(v)}_{xy}, 
	\\ \nt
\mathcal{C}_{\parallel} 
	& \approx \frac{\alpha_{\textrm{R}}}{4\pi} \Gamma^{(v)}_{xx}, 
}
and we have 
\eq{
\Gamma^{(s)} 
	& = {\frac{\Sigma_0}{4\pi }} \int_0^\infty   
	 	\frac{1}{E_{\xi}^2 + (\Sigma'_0 - \gamma \Sigma_z)^2} 
		\lf[1 -  \frac{\beta_{\textrm{I}}^2 + \Sigma_z^2 }{2 u_{\textrm{R}} \xi  + \beta_{\textrm{I}}^2 + \gamma^2 \Sigma_z^2 }   \rf] d\xi
	\approx 		
	{\frac{\Sigma_0}{4(\Sigma'_0 - \gamma \Sigma_z)}} 
	\lf[1 -  \frac{\beta_{\textrm{I}}^2 + \Sigma_z^2 }{2 u_{\textrm{R}} \mu  + \beta_{\textrm{I}}^2 + \gamma^2 \Sigma_z^2 }   \rf]
	\\ 
\Gamma_{xy}^{(v)} & \equiv 
	\int_0^\infty    \frac{\xi \sqrt{ 2u_{\textrm{R}} \xi + \beta_{\textrm{I}}^2}}{2u_{\textrm{R}} \xi  + \beta_{\textrm{I}}^2 + \gamma^2 \Sigma_z^2 } 
	\frac{d\xi}{ E_{\xi}^2 + (\Sigma'_0 - \gamma \Sigma_z)^2}     
	- {\frac{2\pi}{\Sigma_0}}\Gamma^{(s)}
	 \approx 
	\frac{- \pi}{ 2(\Sigma'_0 - \gamma \Sigma_z) }
	\lf[ 1 -  \frac{\beta_{\textrm{I}}^2 + \Sigma_z^2 + \mu \sqrt{ 2u_{\textrm{R}} \mu + \beta_{\textrm{I}}^2}  }{2 u_{\textrm{R}} \mu  + \beta_{\textrm{I}}^2 + \gamma^2 \Sigma_z^2 }   \rf]
	\\
\Gamma_{xx}^{(v)} & \equiv
	\int_0^\infty   \frac{\Sigma_z}{ E_{\xi}^2 + (\Sigma'_0 -  \gamma \Sigma_z)^2}
		\frac{\xi  - \sqrt{2u_{\textrm{R}} \xi + \beta_{\textrm{I}}^2} + \gamma\beta_{\textrm{I}}}
			{ 2u_{\textrm{R}} \xi   + \beta_{\textrm{I}}^2 + \gamma^2 \Sigma_z^2} d\xi 
	\approx 
	\frac{\pi \Sigma_z}{ 2(\Sigma'_0 -  \gamma \Sigma_z)}
		\frac{\mu  - \sqrt{2u_{\textrm{R}} \mu + \beta_{\textrm{I}}^2} + \gamma\beta_{\textrm{I}}}
			{ 2u_{\textrm{R}} \mu   + \beta_{\textrm{I}}^2 + \gamma^2 \Sigma_z^2}. 
}
In particular, in the limit of $ \Sigma_z \ll \beta_{\textrm{I}}$, 
$\mathcal{C}_{\parallel} $ can be represented by 
\eq{
\mathcal{C}_{\parallel} 
	& \approx \frac{\alpha_{\textrm{R}}}{8} 
		\frac{\Sigma_z}{ \Sigma'_0 -  \gamma \Sigma_z}
		\frac{\mu  - \sqrt{ \alpha^2_{\textrm{R}} k_{\textrm{F}}^2 + \beta_{\textrm{I}}^2} + \gamma\beta_{\textrm{I}}}
			{ \alpha^2_{\textrm{R}} k_{\textrm{F}}^2   + \beta_{\textrm{I}}^2 }
	 \approx \frac{\alpha_{\textrm{R}}}{8} 
		\frac{\Sigma_z}{ \Sigma'_0 -  \gamma \Sigma_z}
		\frac{\mu  - \sqrt{ \alpha^2_{\textrm{R}} k_{\textrm{F}}^2 + \beta_{\textrm{I}}^2} + \gamma\beta_{\textrm{I}}}
			{ \alpha^2_{\textrm{R}} k_{\textrm{F}}^2   + \beta_{\textrm{I}}^2 }
}
with 
\eq{ 
& \Sigma'_0  \approx  \biggl[ \frac{1}{2} + \frac{\frac{m \alpha^2_{\textrm{R}}}{\hbar^2}}{2 \sqrt{ \alpha^2_{\textrm{R}} k_{\textrm{F}}^2 + \beta_{\textrm{I}}^2   }} \biggr] \Sigma_0, &
& \Sigma_z  \approx  \frac{\beta_{\textrm{I}}}{2\sqrt{\alpha^2_{\textrm{R}} k_{\textrm{F}}^2 + \beta_{\textrm{I}}^2  }} \Sigma_0, & 
& \gamma \approx \frac{ \beta_{\textrm{I}} }{ \sqrt{\alpha^2_{\textrm{R}} k_{\textrm{F}}^2 +\beta_{\textrm{I}}^2 }}.&
}
where we have used the Fermi wavenumber $k_{\textrm{F}} = \sqrt{ \frac{2m\mu}{\hbar^2}}$, the Rashba energy $u_{\textrm{R}}=\frac{m \alpha^2_{\textrm{R}}}{\hbar^2}$, 
and $\Sigma_z = \frac{\beta_{\textrm{I}}}{2\sqrt{\lambda_0}} \Sigma_0$.

\section{Derivation of $\mathcal{C}_{\parallel}$ and $\mathcal{C}_{\perp}$ in $|\mu-\Sigma_0| > \beta_{\textrm{I}}$ }\label{sec:I-REE_main}
\subsection{Detail of the calculation of $\mathcal{C}_\parallel$ and $\mathcal{C}_\perp$ }
From the same way, we obtain the spin density as   
\eq{ \nt  
\langle \bm{s}^{\textrm{VEE},i}_{\textrm{v}} \rangle 	
	& = -\frac{e\hbar}{2\pi} \Gamma^{(v)}_{jn} 
		\lf[ \delta_{ni} + \Gamma^{(s)}_{ni} + \Gamma^{(s)}_{n\ell_1}\Gamma^{(s)}_{\ell_1 i} +\cdots
		\rf] E_{j}
	\\ \label{eq:1-2-b} 
& = e \nu_e \left[\mathcal{C}_\perp (\hat{\bm{z}}\times \bm{E})_i +  \textrm{v}  \mathcal{C}_\parallel E_i  \right]
}
with 
\eq{
\mathcal{C}_{\perp} 
	& = \frac{\alpha_{\textrm{R}} }{4\pi}  \Gamma^{(v)} 
		\frac{ 1- \Gamma^{(s)}_{xx}}{ [1-\Gamma^{(s)}_{xx}]^2 + [\Gamma^{(s)}_{xy}]^2}   
	\\ 
\mathcal{C}_{\parallel} 
	& = \frac{\alpha_{\textrm{R}} }{4\pi } \Gamma^{(v)} \frac{\Gamma^{(s)}_{xy}}{ [1-\Gamma^{(s)}_{xx}]^2 + [\Gamma^{(s)}_{xy}]^2}	   
	\\
\label{eq:S_xx}
\Gamma_{xx}^{(s)} = \Gamma_{yy}^{(s)}  
	&	= \frac{\Sigma_0}{\pi} \int_0^\infty d\xi
		\frac{ (\xi-\mu)^2 + \Sigma_0^2 - \beta_{\textrm{I}}^2 }{ ( E_+^2 +  \Sigma_0^2)  ( E_-^2 +  \Sigma_0^2)}
	\\  \label{eq:S_xy}
\Gamma_{xy}^{(s)} = -\Gamma_{yx}^{(s)} 
	& 	=   \frac{\Sigma_0}{\pi} \int_0^\infty d\xi
			 \frac{ 4u_z \Sigma_0 }{(E_+^2 +\Sigma_0^2)(E_-^2 +\Sigma_0^2)} 
	\\
\label{eq:V_xy}
\Gamma^{(v)}
	 & = 4\int_0^\infty d\xi 
		 \frac{\xi - \mu- \frac{u_{\textrm{R}} }{u^2} \lf[ (\xi - \mu)^2 + \Sigma_0^2 \rf] 
		   }{ (E_+^2 +  \Sigma_0^2)(E_-^2 +\Sigma_0^2)  }\xi. 
}
where we have used $\Sigma_0 = \pi \nu_e n_{\textrm{c}} u_{\textrm{i}}^2$, $\alpha_{\textrm{R}}^2 k^2 = 2u_{\textrm{R}}\xi$, $u_{\textrm{R}}=m\alpha^2_{\textrm{R}}/\hbar^2$, $E_{\pm}  = \xi - \mu  \pm u$, and $u= \sqrt{2 u_{\textrm{R}} \xi + \beta_{\textrm{I}}^2}$.
Here, $\mathcal{C}_{\perp} $ and $\mathcal{C}_{\parallel} $ in the above equations are approximately given within $u_{\textrm{R}}\ll \Sigma_0$ and $\mu \gg \Sigma_0$ as 
\eq{
\mathcal{C}_{\perp} 
	& \approx \frac{\alpha_{\textrm{R}} }{4\pi}  \Gamma^{(v)}  
		\approx \frac{ \alpha_{\textrm{R}}  }{ 2\Sigma_0} \lf[ 1 - \frac{1}{2} \frac{\alpha_{\textrm{R}}^2 k_{\textrm{F}}^2 }{ \beta_{\textrm{I}}^2 + \alpha_{\textrm{R}}^2 k_{\textrm{F}}^2 } \rf] 
	\\ 
	\label{eq:c-parallel-app1}
\mathcal{C}_{\parallel} 
	& \approx \frac{\alpha_{\textrm{R}} }{4\pi } \Gamma^{(v)} \Gamma^{(s)}_{xy}
	\approx  
	 \frac{ \alpha_{\textrm{R}}\beta_{\textrm{I}} }
		  { \beta_{\textrm{I}}^2 + \alpha_{\textrm{R}}^2 k_{\textrm{F}}^2 } 
	 \lf[ 1 - \frac{1}{2} \frac{\alpha_{\textrm{R}}^2 k_{\textrm{F}}^2 }{ \beta_{\textrm{I}}^2 + \alpha_{\textrm{R}}^2 k_{\textrm{F}}^2 } \rf] 	  
}
where we have used  $|\Gamma^{(s)}_{xx}]|<1$, $|\Gamma^{(s)}_{xy}|<1$, and the following results: 
\eq{  \label{eq:S_xy-2}
\Gamma_{xy}^{(s)} 
	& 	\approx 
		 \frac{ 2 \beta_{\textrm{I}} \Sigma_0}{ \beta_{\textrm{I}}^2 + \alpha_{\textrm{R}}^2 k_{\textrm{F}}^2   +\Sigma_0^2 }
		 \approx
		  \frac{ 2 \beta_{\textrm{I}} \Sigma_0}
		  { \beta_{\textrm{I}}^2 + \alpha_{\textrm{R}}^2 k_{\textrm{F}}^2 } \ \ \ \ ( \mu  \gg \Sigma_0 )
	\\
\label{eq:V_xy-2}
\Gamma^{(v)}
	 & \approx 	
		\frac{ 2\pi  }{ \Sigma_0} \lf[ 1 - \frac{1}{2} \lf(1-  \frac{\beta_{\textrm{I}}^2}{ \beta_{\textrm{I}}^2 + \alpha_{\textrm{R}}^2 k_{\textrm{F}}^2 }\rf)	 \rf] 
		=
		\frac{ 2\pi }{ \Sigma_0} \lf[ 1 - \frac{1}{2} \frac{\alpha_{\textrm{R}}^2 k_{\textrm{F}}^2 }{ \beta_{\textrm{I}}^2 + \alpha_{\textrm{R}}^2 k_{\textrm{F}}^2 } \rf].
		 \ \ \ \ ( \mu  \gg \Sigma_0 )
}
In the above equations, we also used the following integrals: 
\eq{
\int_0^\infty  \frac{d\xi  }{ (E_+^2 +  \Sigma_0^2)(E_-^2 +\Sigma_0^2)  }
	& \approx 
	 \frac{\pi}{2\Sigma_0 (\beta_{\textrm{I}}^2 + \alpha_{\textrm{R}}^2 k_{\textrm{F}}^2 +2u_{\textrm{R}}^2 +\Sigma_0^2  )} 
	 \approx
	  \frac{\pi}{2\Sigma_0 (\beta_{\textrm{I}}^2 + \alpha_{\textrm{R}}^2 k_{\textrm{F}}^2  )} 
	\\
\int_0^\infty d\xi 
		 \frac{  4(\xi - \mu)\xi }
		        { (E_+^2 +  \Sigma_0^2)(E_-^2 +\Sigma_0^2)  }	
		        & \approx \frac{2\pi}{\Sigma_0} 
	\\
4\int_0^\infty d\xi 
		 \frac{  \frac{u_{\textrm{R}} \xi }{2u_{\textrm{R}} \xi + \beta_{\textrm{I}}^2 } [(\xi -\mu)^2 + \Sigma_0^2] }
		        { (E_+^2 +  \Sigma_0^2)(E_-^2 +\Sigma_0^2)  }	
		        & = 2 \int_0^\infty d\xi 
		 \frac{  (\xi - \mu)^2 + \Sigma_0^2 }
		        { (E_+^2 +  \Sigma_0^2)(E_-^2 +\Sigma_0^2)  }	
		        - 
		        \int_0^\infty d\xi 
		 \frac{ \frac{2\beta_{\textrm{I}}^2  }{2u_{\textrm{R}} \xi + \beta_{\textrm{I}}^2 } [ (\xi - \mu)^2 + \Sigma_0^2 ]}
		        { (E_+^2 +  \Sigma_0^2)(E_-^2 +\Sigma_0^2)  }
	\\
2 \int_0^\infty d\xi 
		 \frac{  (\xi - \mu)^2 + \Sigma_0^2 }
		        { (E_+^2 +  \Sigma_0^2)(E_-^2 +\Sigma_0^2)  }	
		        & \approx \frac{\pi}{\Sigma_0} 
		        \lf[ 1 + \frac{\Sigma_0^2}{\beta_{\textrm{I}}^2 + \alpha_{\textrm{R}}^2 k_{\textrm{F}}^2 + \Sigma_0^2 }\rf] 
	\\
\int_0^\infty d\xi 
		 \frac{ \frac{2\beta_{\textrm{I}}^2  }{2u_{\textrm{R}} \xi + \beta_{\textrm{I}}^2 } [ (\xi - \mu)^2 + \Sigma_0^2 ]}
		        { (E_+^2 +  \Sigma_0^2)(E_-^2 +\Sigma_0^2)  }
		& \approx 
		\frac{\pi}{\Sigma_0}  	\lf[ 1 + \frac{\Sigma_0^2}{\beta_{\textrm{I}}^2 + \alpha_{\textrm{R}}^2 k_{\textrm{F}}^2 + \Sigma_0^2 } \rf]	 \frac{\beta_{\textrm{I}}^2}{ \beta_{\textrm{I}}^2 + \alpha_{\textrm{R}}^2 k_{\textrm{F}}^2 }.
}
Here, $k_{\textrm{F}} \equiv \frac{2m\mu}{\hbar^2}$ is the Fermi wave number.
\begin{figure}[ht]
\includegraphics[width=15cm]{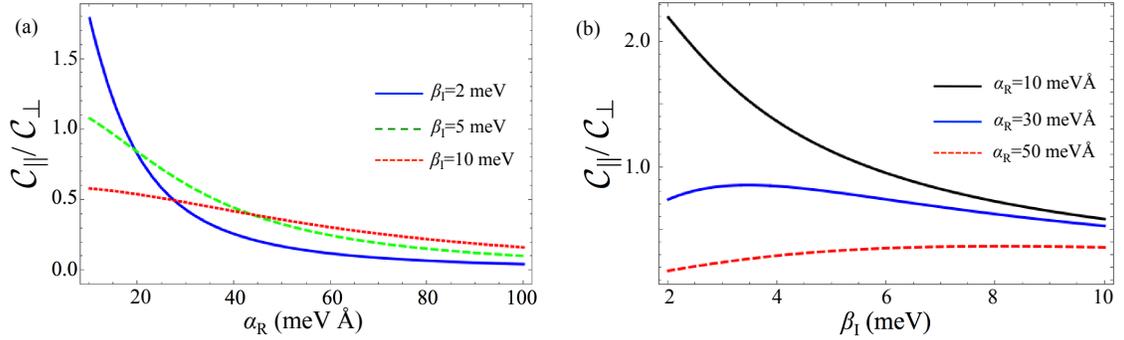}
\caption{
(a) The Rashba SOC ($\alpha_{\textrm{R}}$) dependence of $\mathcal{C}_\parallel /\mathcal{C}_\perp $ for several the Ising SOC ($\beta_{\textrm{I}}$) in the conduction band in $\mu > \beta_\textrm{I} $.
(b) $\beta_\textrm{I}$ dependence of $\mathcal{C}_\parallel /\mathcal{C}_\perp $  for several Rashba SOC ($\alpha_{\textrm{R}}$). In these figures, we used $\Sigma_0 =3$ meV and $\mu=0.1$ eV.
We find that $\mathcal{C}_\parallel $ can be comparable to $\mathcal{C}_\perp$ in a realistic parameter regime.
}
\label{figS3} 
\end{figure} 
Figure S\ref{figS3}(a) [(b)] shows the $\alpha_{\textrm{R}}$ [$\beta_{\textrm{I}}$] dependence of $\mathcal{C}_{\parallel}/\mathcal{C}_{\perp}$ for several $\beta_{\textrm{I}}$ [$\alpha_{\textrm{R}}$] in $\mu>\beta_{\textrm{I}}$.
We find that $\mathcal{C}_{\parallel}$ can be comparable to $\mathcal{C}_{\perp}$.

\subsection{Relation between $\mathcal{C}_\parallel$ and the Berry curvature in the gated TMD} 
Interestingly, $\mathcal{C}_\parallel$ can be also represented by using Berry curvature in the gated MTMDs. The Berry curvature in the presence of the Rashba and Ising SOCs at each valley are given by 
\eq{
\Omega^{\textrm{v}=\pm}_{\textrm{spin}} (k_\textrm{F})
	& = \hat{z}\cdot\bm{\nabla} \times \langle \bm{k},\textrm{v} | i\bm{\nabla} |\bm{k}, \textrm{v} \rangle|_{|\bm k|=k_\textrm{F}}  
	= \textrm{v} \frac{\alpha_{\textrm{R}}^2 \beta_{\textrm{I}} }{2( \alpha_{\textrm{R}}^2 k_{\textrm{F}}^2  + \beta_{\textrm{I}}^2 )^{3/2}} .
} 

By using the Berry curvature of the gated TMD $\Omega_{\textrm{spin}}^{\textrm{v}=\pm}$, $\mathcal{C}_{\perp} $ is represented from Eq. (\ref{eq:c-parallel-app1}) as 
\eq{ \nt 
\mathcal{C}_{\parallel} 
	& \approx  
	 \frac{ \alpha_{\textrm{R}}^2  \beta_{\textrm{I}} }
		  { (\alpha_{\textrm{R}}^2 k_{\textrm{F}}^2 + \beta_{\textrm{I}}^2 )^{3/2} } 
		   \frac{1}{\alpha_{\textrm{R}}} \frac{ \alpha_{\textrm{R}}^2 k_{\textrm{F}}^2  + 2\beta_{\textrm{I}}^2  }
		  { \sqrt{\alpha_{\textrm{R}}^2 k_{\textrm{F}}^2  + \beta_{\textrm{I}}^2} } 
	\\ \nt 
	& = |\Omega_{\textrm{spin}}^{\textrm{v}} (k_{\textrm{F}})  | k_{\textrm{F}}
		   \frac{ \alpha_{\textrm{R}} k_{\textrm{F}} }{ \sqrt{\alpha_{\textrm{R}}^2 k_{\textrm{F}}^2  + \beta_{\textrm{I}}^2} } 
		  \biggl[ 1+ \frac{2\beta_{\textrm{I}}^2}{\alpha_{\textrm{R}}^2 k_{\textrm{F}}^2} \biggr]
	\\
	& = |\Omega_{\textrm{spin}}^{\textrm{v}} (k_{\textrm{F}})  | k_{\textrm{F}}
		 \cos{\theta_{\parallel}}
		  \biggl[ 1+ \frac{2\beta_{\textrm{I}}^2}{\alpha_{\textrm{R}}^2 k_{\textrm{F}}^2} \biggr]		  
}
where we define $ \cos{\theta_{\parallel}} \equiv  \frac{ \alpha_{\textrm{R}} k_{\textrm{F}} }{ \sqrt{\alpha_{\textrm{R}}^2 k_{\textrm{F}}^2  + \beta_{\textrm{I}}^2} } $.

\section{Detection of VEE using Kerr effect measurements}\label{sec:detection_way}
In this section, we provide details of the detection scheme of VEE using Kerr effect measurements. 
To detect the in-plane parallel magnetization due to the VEE, we consider the longitudinal Kerr effect, as described in Fig. 5 in the main text.
After applying the in-plane electric field ($E_x$), spin density can be generated by VEEs in the steady state, which induces nonzero magnetization in gated MTDMs. By focusing a beam of laser with $\textrm{s}(\textrm{p})$-polarizations onto the system, the non-zero magnetization couples differentially with left-handed and right-handed components, which leads to a superposition of both $\textrm{s}$- and $\textrm{p}$-polarized lights in the reflected beam. The reflection coefficients $r_{ij}$ for $i,j=s, p$-polarized lights are given by\cite{You1998} 
\eq{ \label{eq:reflection-pp}
r_{pp} & = \frac{n_2 \cos{\theta_0} - n_0 \cos{\theta_2} }
		       {n_2 \cos{\theta_0} + n_0 \cos{\theta_2} }
	     -  \frac{ 4\pi i n_0 d_1 \cos{\theta_0} (n_2^2 \cos{\theta_1}^2 - n_1^2 \cos{\theta_2}^2) }
		       {\lambda (n_0 \cos{\theta_2} + n_2 \cos{\theta_0})^2 } , 		       
	\\
	\label{eq:reflection-sp}
r_{sp} & = \frac{4\pi n_0 n_1 Q d_1 \cos{\theta_0} (M_z n_1 \cos\theta_2 + M_x n_2 \sin\theta_1)}
		       { M_{tot} (n_0 \cos{\theta_0} + n_2 \cos{\theta_2} ) (n_0 \cos{\theta_2} + n_2 \cos{\theta_0}) }
	\\
	\label{eq:reflection-ps}
r_{ps} & = \frac{4\pi n_0 n_1 Q d_1 \cos{\theta_0} (M_z n_1 \cos\theta_2 - M_x n_2 \sin\theta_1)}
		       { M_{tot} (n_0 \cos{\theta_0} + n_2 \cos{\theta_2} ) (n_0 \cos{\theta_2} + n_2 \cos{\theta_0}) }, 
}
where $n_l$ and $\theta_{l}$ ($l=0, 1, 2$) denotes the refractive index and the incident angle at the $l$-th medium (shown in Figure S \ref{fig:Kerr_effect_mesurement}).
$Q$ is the Voigt vector depending on materials, 
$d_1$ is the thickness of the magnetic medium, 
$\lambda$ is wavelength of the light, 
$M_i$ ($i=x,y,z$) are the $i$-component of the magnetization, 
and $M_{tot}=\sqrt{M_x^2+M_y^2+M_z^2}$ is the magnitude of  the magnetization. 
\begin{figure}[tb]
\includegraphics[width=8cm]{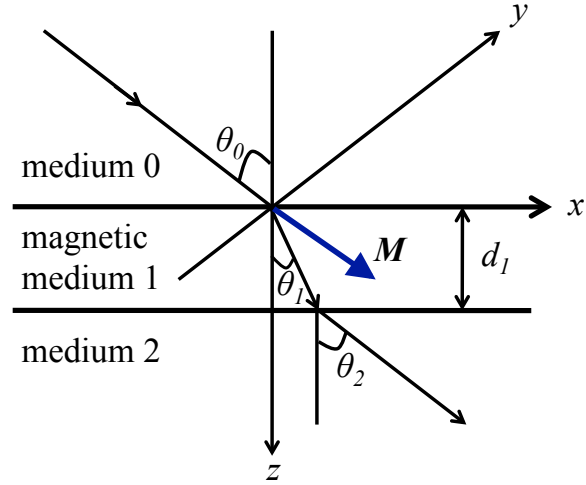}
\caption{
The coordinate system for the Kerr effect measurement in the nonmagnetic medium 0 and 2 and the magnetic medium 1, where the magnetization (blue arrow) is polarized along arbitrary direction. 
}
\label{fig:Kerr_effect_mesurement} 
\end{figure} 
From Eqs. (\ref{eq:reflection-pp})-(\ref{eq:reflection-ps}), the Kerr angle for $i=s, p$-polarized light, $\theta_{\textrm{K}}^{i}$, is given by\cite{You1998}
\eq{
\theta_{\textrm{K}}^{p} 
	& = \frac{\cos{\theta_0} }{\cos{(\theta_0+\theta_2) }}
	\lf( \frac{M_x}{M_{tot}} \frac{\sin{\theta_1}^2 }{\sin{\theta_2}}
		+ \frac{M_z}{M_{tot}} \cos{\theta_2}  \rf)\Theta_{n}, 
	\\		
\theta_{\textrm{K}}^{s} 
	& = \frac{\cos{\theta_0} }{\cos{(\theta_0-\theta_2) }}
	\lf( \frac{M_x}{M_{tot}} \frac{\sin{\theta_1}^2 }{\sin{\theta_2}}
		- \frac{M_z}{M_{tot}} \cos{\theta_2}  \rf)\Theta_{n}. 
}
Here, $\Theta_{n} $ is defined as the complex polar Kerr effect for normal incidence in the film given by 
\eq{
\Theta_{n} & \equiv \frac{4\pi n_0 n_1^2 Q d}{\lambda (n_s^2 - n_0^2 )}
} 
The Voigt vector $Q$ is determined by the the Kerr angle for the $p$-polarized wave under the normal incident light $(\theta_{\textrm{K}}^{p})^{\textrm{normal}} $ as 
\eq{
(\theta_{\textrm{K}}^{p})^{\textrm{normal}}
	& \equiv \frac{r_{sp} (\theta_0 = 0)}{r_{pp} (\theta_0 = 0)}
		= \frac{ 4\pi n_0 n_1^2 Q d_1 \cos{\theta_0} \cos{\theta_2} }
		       { (n_0 \cos{\theta_0} + n_1 \cos{\theta_2})(n_1 \cos{\theta_0} - n_0 \cos{\theta_2}) }.
}

By considering medium 0 and medium 2 with similar refractive indices (i.e., $n_0 \approx n_2$ and $\theta_0 \approx \theta_2$ ), the relation between the in-plane magnetization $M_x$ and Kerr angle is given by
\eq{
M_x & = \frac{1}{2\Theta_n} \frac{\sin{\theta_0} \cos{\theta_0}  }{\sin{\theta_1}^2}
	\lf[ \theta_{\textrm{K}}^{s} + \theta_{\textrm{K}}^{p} \cos{(2\theta_0)}  \rf]  M_{tot}
}
Based on the relation above, for oblique incidence with the incident angle $\theta_0 \approx \pi/4$, we have 
\eq{
M_x \propto \theta_{\textrm{K}}^{s}.}
as discussed in the main text. Therefore, the parallel spin density due to VEEs can be mapped out by $\theta_{\textrm{K}}^{s}$.

\section{The induced spin density}\label{sec:spin-density}
Finally, we estimate the induced spin density. 
The spin density of each valley can be estimated as $s_\perp  = e \nu_e \mathcal{C}_{\perp} |\bm{E}\times \hat{\bm{z}}| $ and $s_\parallel  = e \nu_e \mathcal{C}_{\parallel} |\bm{E}| $, respectively. 
These are estimated as $s_\perp \approx 13\ \mu$m$^{-2}$ and $s_\parallel \approx 8 \ \mu$m$^{-2}$ respectively,
when we apply a dc electric field $E_x =100$ mV/$\mu$m for the system with $\beta_{\textrm{I}} = 10$ meV, $\alpha_{\textrm{R}} = 10$ meV\AA, $\mu=0.1$ eV, $\Sigma_0=3$ meV, and $m/m_e = 0.5$ with $m_e$ being the electron rest mass. The conventional Edelstein effect in the TMD material can be measured because the observed spin density in the interface of the InGaAs/GaAs  \cite{Kato2004} is only about $\rho_{\textrm{el}} d \approx 8\ \mu$m$^{-2}$ under the thickness of the film $d\approx 1\ \mu$m \cite{Kato2004-2}.

\end{widetext}
\end{document}